\documentclass{aa}  

\usepackage{graphicx}
\usepackage{txfonts}
\usepackage{color}
\usepackage{soul}
\usepackage{natbib}
\begin{document} 

%-----------------------------------------------------------------------
\title{
Prestellar core modeling in the presence of a filament
      }
\subtitle{
The dense heart of L1689B
      }

\author{
J. Steinacker
\inst{1,2,3}
\and
A. Bacmann
\inst{1,2}
\and
Th. Henning
\inst{3}
\and
S. Heigl
\inst{4}
       }

\institute{
Univ. Grenoble Alpes, IPAG, F-38000 Grenoble, France\\
\email{stein@mpia.de}
\and
CNRS, IPAG, F-38000 Grenoble, France
\and
Max-Planck-Institut f\"ur Astronomie,
K\"onigstuhl 17, D-69117 Heidelberg, Germany
\and
University Observatory Munich, LMU Munich, 
Scheinerstr. 1, D-81679 Munich, Germany
          }

\date{Received; accepted}

%-----------------------------------------------------------------------
\abstract
% context 
{
Lacking a paradigm for the onset of star formation, it is important to
derive basic physical properties of prestellar cores and filaments like density 
and temperature structures.
}
% aims
{
We aim to disentangle the spatial variation in density
and temperature across the prestellar core L1689B,
which is embedded in a filament. 
We want to determine the range of possible central densities and 
temperatures that are consistent with the continuum radiation data.
}
% methods
{
We apply a new synergetic radiative transfer method: 
the derived 1D density profiles are both consistent with a cut through
the {\it Herschel} PACS/SPIRE and JCMT SCUBA-2 continuum 
maps of L1689B and with a derived local interstellar radiation field.
Choosing an appropriate cut along the filament major axis, we minimize 
the impact of the filament emission on the modeling.
}
% results
{
For the bulk of the core (5000-20000 au) an isothermal sphere model 
with a temperature of around 10 K provides the best fits. 
We show that the power law index of the density profile, as well as the constant 
temperature can be derived directly from the radial surface brightness 
profiles.
For the inner region (< 5000 au), we find a range of densities and 
temperatures that are consistent with the surface brightness profiles and 
the local interstellar radiation field.
Based on our core models, we 
find that pixel-by-pixel single temperature spectral energy distribution fits are incapable of
determining dense core properties.
}
% conclusions
{
We conclude that, to derive physical core properties, it is important to avoid
azimuthally-averaging core and filament.
Correspondingly, derived core masses are too high since they include 
some mass of the filament, and might introduce errors when determining
core mass functions. 
The forward radiative transfer methods also avoids the loss of information owing
to smearing of all maps to 
the coarsest spatial resolution.
We find the central core region to be colder and denser than estimated in 
recent inverse radiative transfer modeling, possibly indicating the start of
star formation in L1689B.
}


\maketitle
 
%+++++++++++++++++++++++++++++++++++++++++++++++++++++++++++++++++++++++
\section{Introduction}
For decades, no global answer has been known to the question of what ends the life
of a prestellar low-mass core and starts the star formation process
\citep{2007ARA&A..45..339B}.
Is it the weakening of a supporting force based on magnetic fields or 
turbulence, is it an internal or external kinetic trigger, or a mass 
overload? 
As recent data, e.g., from {\it Herschel}, better probe the complex 
filamentary network from which cores emerge, attention also turns to 
the role of the hosting filaments
\citep{2014prpl.conf...27A}.

To find the physical reason of the star formation onset, substantial 
effort has been put in determining the basic properties of prestellar 
cores and filaments 
\citep{2016MNRAS.tmp..350M}, that is the gas density and temperature structures
\citep{2014A&A...569A...7S}
and the topology of the velocity field.
\citep{2011A&A...533A..34H,2013A&A...554A..55H}.

Tracing the gas physical properties directly by the line emission of molecules is
difficult since their abundances and excitation conditions vary with the physical 
properties so that different molecular emission lines trace different regions.
Unfortunately, their abundances are also reduced by a freeze-out onto 
the dust grains because of the low temperatures around 10 K.
Therefore, hardly any molecular tracers are available
 in the central region where the star formation process is initiated
 \citep{2013A&A...559A..53M}.
Hence, the gases' physical structure is determined from the radiation 
emitted or scattered by 
small dust grains that are mixed in the gas
\citep[e.g.,][]{2013A&A...551A..98L}, 
and that also have the
same temperature as the gas above a certain density threshold. 

One of the "standard" prestellar cores that was investigated in detail 
both by analyzing its line emission and by dust radiative transfer (RT)
effects is the core B68
\citep[e.g.,][]{2001Natur.409..159A}.
The most complete modeling of the core was performed by 
\citet{2012A&A...547A..11N}
based on dedicated {\it Herschel} far-infrared (FIR) continuum maps
and ancillary sub-mm continuum maps
(we refer to this paper for a list of references for work
on B68). 
A particularly interesting result of their
ray-tracing analysis was the steep slope of the radial density profile 
($\propto r^{-3.5}$ azimuthally averaged for radii $r>60"$ with 
density deviations
up to a factor of 4 in certain directions). 
Prestellar cores are expected to
have density profiles resembling that of an isothermal sphere 
\citep{1977ApJ...214..488S} or of a
Bonnor-Ebert sphere (BES)
\citep[e.g.,][]{1996ApJ...469..194M},
which are characterized by a decrease $\propto r^{-2}$ in the outer parts.
Magnetized simple filament models have steeper gradients perpendicular to the
filament axis $\propto r^{-4}$
\citep{1964ApJ...140.1056O}. However, for weak magnetization,
\citet{2003ApJ...593..426T} found $\propto r^{-2}$.
For the general problem of collapsing cylinders see also \citet{2003A&A...411....9H}.
More recent theoretical work
\citep[e.g.,][]{2012A&A...542A..77F,2014MNRAS.445.2900S} suggests flatter
profiles closer to an $r^{-2}$-decline, in agreement with observational
studies
\citep[e.g.,][]{2013A&A...550A..38P}.
Because its environment and that of the neighboring cores only show weak remnants of the hosting filament, and its outer density profile is steeper than expected for a Bonnor-Ebert sphere model, B68 might not be representative of most prestellar cores.
According to 
\citet{2014prpl.conf...27A},
70\% of all cores identified in the 
{\it Herschel} Gould Belt Survey (HGBS) are located in a filament.
Hence, it appears that the common mode of the onset of star formation
may be a core that is still embedded and potentially affected by the hosting filament.

In this work, we therefore analyze the well-known core-filament system L1689B in the Ophiuchus star-forming region. 
The central part of the core shows depletion of gaseous molecules 
\citep[e.g., CO,][]{2001MNRAS.323.1025J},
but the level of depletion is moderate so that 
the core is either young, or has evolved quickly towards high degrees
of concentration.

The core presents line asymmetries seen in e.g.  CS(2$-$1), 
H$_2$CO(2$_{12} -1_{11}$) and HCO$^+$(3$-$2), which are 
commonly interpreted as indication of infall motions
\citep{2000A&A...361..555B, 2000ApJ...538..260G, 2001ApJS..136..703L}.
However, the projection along the line-of-sight and
limited spatial resolution
make the interpretation difficult. Additional velocity components have been 
proposed:
L1689B may show oscillations as suggested for some cores by 
\citet{2003ApJ...586..286L}
and 
\citet{2010ApJ...721..493B}.
\citet{2004MNRAS.352.1365R}
argued that a rotating, non-infalling core center with a size of 3000 au may 
account for the shape of the line profiles seen in HCO$^+$.
\citet{2013ApJ...769...50S}
considered the effect of internal velocities on the stability of BES.
Analyzing L1689B, they concluded that the infall motion hints towards a kinematic
disturbance as the motions are larger than expected for spontaneous gravitational infall.

\citet{2013ApJ...777..121S}
compared the degree of deuterium fractionation with infall velocities
calculated from the HCO$^+$ spectra. 
They find that the Jeans stability is a useful indicator
of collapse. 
Calculating the Jeans mass, however, relies on the knowledge of
the temperature and the outer core radius which seems to be unclear for L1689B:
the derived masses and central column density of the core from the literature span a surprisingly wide 
range, with non-overlapping uncertainties.
According to 
\cite{2005MNRAS.360.1506K}
and 
\citet{2016MNRAS.458.2150M}
L1689B is a low-mass core with
0.4$\pm$0.1 M$_\odot$ and 0.49$\pm$0.05 M$_\odot$, respectively.
Contrary, \citet{2000A&A...361..555B} and
\cite{2014A&A...562A.138R} (R+14 hereafter) give a core mass of
11 \,M$_\odot$ and
11$\pm$2\,M$_\odot$, respectively, about a factor of 20 larger than in 
\cite{2005MNRAS.360.1506K}.

The central column densities for H$_2$ were derived as 
1.4$\pm$0.1 $\times$10$^{26}$ m$^{-2}$,
3.6$\pm$0.1 $\times$10$^{26}$ m$^{-2}$, and 
5$\times$10$^{26}$ m$^{-2}$ by
\citet{2016MNRAS.458.2150M}, R+14, and \citet{2005MNRAS.360.1506K},
respectively.
Concerning the central temperature, the three publications give dust temperatures
in the range 9.3-13 K (10.9$\pm$0.2 K, 9.8$\pm$0.5 K, and 11$\pm$2 K, respectively).
The distance estimates range from 120 pc
\citep{2008ApJ...675L..29L, 2008A&A...480..785L}
over 140 pc
(R+14)
to 165 pc
\citep{2005ApJ...619..379C}.
In this work, we use 140 pc for the sake of comparison with
R+14.

From these results, L1689B seems to be a core that already undergoes collapse although
it is still chemically young. 
However, the rather low central densities derived in 
the aforementioned studies seem difficult to reconcile with the fact that 
the core is already collapsing.
The density structure should show a strong central increase
as it is seen for the core L1544 and expected for a contracting BES
\citep{2010MNRAS.402.1625K}.
Along this way, three difficulties arise.

(1) Is dust continuum modeling of single-dish telescope data able
to reveal a central density increase on length scales of a few 1000 au (henceforth,
we use the abbreviation kau=1000 au)?
R+14 have applied an Abel transform technique to L1689B (and B68) 
deriving density and dust temperature profiles from {\it Herschel} dust emission maps. 
It requires, however, to smear the continuum maps at the different
wavelengths to a single (coarse) resolution which decreases the sensitivity to 
central gradients.

(2) How is the mass of the core determined when the core is located within a 
filament and a spherically or elliptically symmetric model mixes surface
brightness contributions from core and filament in the maps?

(3) 
Because of the strong
shielding of external radiation due to the steep density increase in the core center, the core temperature decreases
towards the center.
Is the decreased efficiency of dust thermal emission at low temperatures hindering the
detection of the central density enhancements?
\citet{2015A&A...574L...5P} stressed this point
concerning a potential reservoir of central cold dust (< 8 K) that could be missed in the 
{\it Herschel} data modeling leading to errors in the derived total core mass.
They argued that the ambiguity in dust mass estimates from FIR/mm data alone 
give rise to mass uncertainties to up to 70\% for L1689B.

In this work, we address these three questions. We
perform forward RT modeling of five FIR and submillimeter continuum maps of 
the core-filament system L1689B ({\it Herschel} PACS/SPIRE, JCMT/SCUBA-2) and additional
RT modeling based on a grid of possible local radiation fields.
In Sect. \ref{Dise} we investigate physical motivations and practical criteria to distinguish
core and filament also based on
hydrodynamical calculations results obtained with the code RAMSES, 
and propose a scheme to extract the core emission from the total emission.
The basic numerical method to derive the core density and temperature is discussed in 
Sect.~\ref{Nume}
and applied in Sect.~\ref{Deri}. 
This section also explains an additional method to remove ambiguities in the modeling by estimating the outer radiation field
from the dust emission in all model shells.
In Sect. \ref{Disc}, we compare our findings with earlier obtained results about the 
interstellar radiation field (ISRF), the mass and column density of L1689B, about determining dense core properties
based on pixel-by-pixel single temperature spectral energy distribution (SED) fitting, the impact of filaments on
the core mass function, the impact of opacity changes, and
the mass of cold dust with temperatures < 8 K in L1689B.

%+++++++++++++++++++++++++++++++++++++++++++++++++++++++++++++++++++++++
\section{Disentangling core and filament} \label{Dise}
There are several motivations to distinguish the prestellar core and filament.
\begin{enumerate}
\item
{\it Symmetry change in the spatial model:} 
With a core showing central symmetry embedded in a filament with expected
axial symmetry, the break-down of the spherical symmetry assumption marks
also a border for the applicability of 1D models.
\item
{\it Stability against gravitational collapse:} 
The simplest way to evaluate the stability of the core is to determine the
Jeans mass which requires to know the core radius that is not well-defined in a 
core-filament system.
\item
{\it Region with gas contributing to the star formation process:} 
Not all of the core and filament gas will end in the star.
Physically interesting is the region where gas and dust that are able to fall onto
the central region before the star has reached its final mass. 
Its border is difficult to determine and likely more complex than estimating
the free-fall radius for a given evolution time.
\item
{\it Region with gas affecting the central temperature:}
Since the central core temperature depends on the incoming radiation,
the outer core and filament matter affects the physics in the star
formation zone. 
The main consequence for low-mass cores is a shielding by outer regions.
Also this border is difficult to find as usually the 3D geometry of the
system is unknown and shielding effects strongly depend on 3D 
sight lines.
\end{enumerate}
%-----------------------------------------------------------------------
\begin{figure}
\vbox{
\includegraphics[width=9cm]{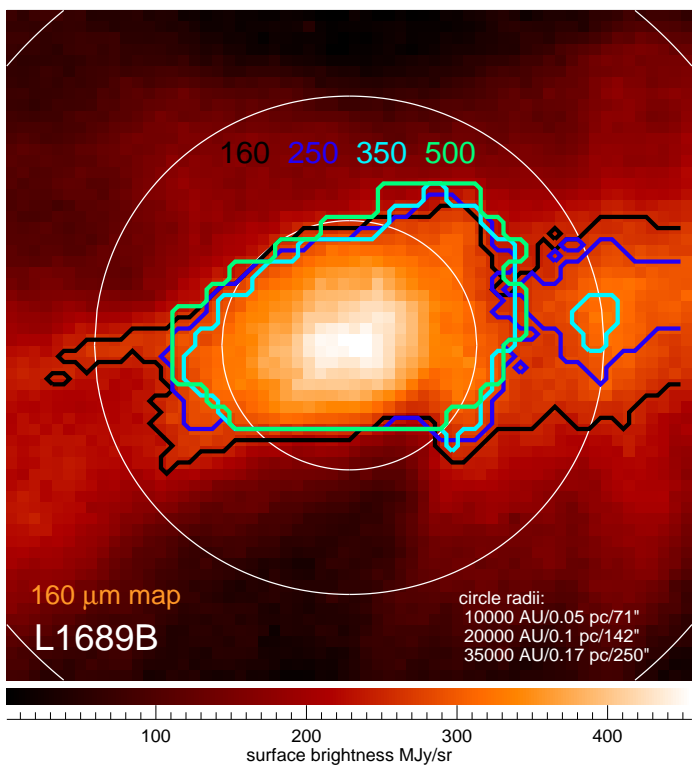}
     }
\caption{
Contours where the intensity of the core reaches the filament maximal
intensity (reference point between the two vertical sub-filaments)
for the different wavelengths. 
The color-coded image shows the 160 $\mu$m map.
        }
\label{comparable}
\end{figure}
%-----------------------------------------------------------------------
\subsection{Core-filament border estimates}\label{borders}
A first method to distinguish core and filament comes from
looking at the maps where the intensity of the core has dropped to 
the mean maximal value of the filament outside the core.
We used the L1689B {\it Herschel} PACS and SPIRE 
\citep{2010A&A...518L...3G}
HGBS \citep{2010A&A...518L.102A} data\footnote{
provided with a detailed description about the observations 
and data reductions at http://gouldbelt-herschel.cea.fr/archives.
}.
Fig.~\ref{comparable} shows the 160 $\mu$m map where the 
filament is seen in most detail and overlaid 
contours where this value is reached. 
As reference points, we chose the point with the maximal filament 
intensity to the
right of the core between the two vertical sub-filaments.
This avoids confusion with the overlapping
sub-filaments to the right but still characterizes the maximal 
intensity expected from the main filament.
Comparing with the white circles at 10, 20, and 30 kau shows that this
value is reached at radii between 8 kau and 15 kau depending on 
the location within the filament. 
We note that the limiting contour radius only weakly depends on
the wavelengths. This means that the radiation we receive from radii
of 20 kau or larger is clearly dominated by the filament.

\begin{figure}
\vbox{
\includegraphics[width=9cm]{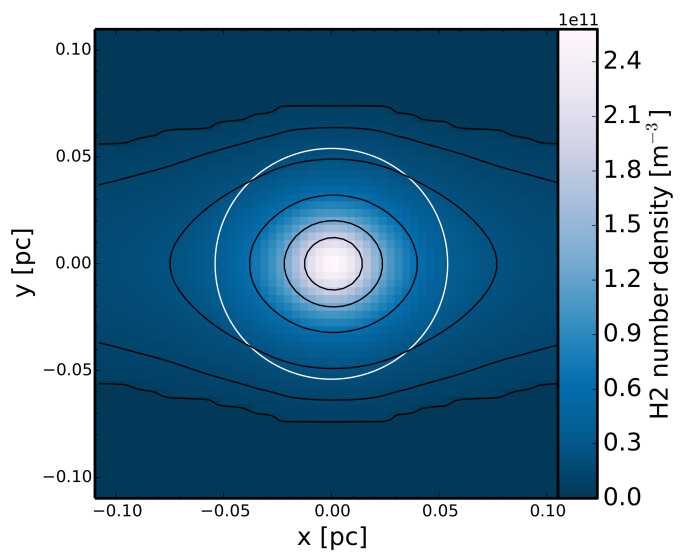}
     }
    \caption{
	    Cut through the H$_2$ number density structure including the core
	    center for a hydro-dynamical simulation of the evolution
	    of a filament with the RAMSES code.
	    The white circle indicates the radius where the density deviation
	    from spherical symmetry exceeds 50\%.
            }
    \label{RAMSESsym}
\end{figure}

More insight into distinguishing core and filament can be gained from a glance at
typical results from hydrodynamical calculations of a core-forming filament
system.
Fig.~\ref{RAMSESsym} shows a cut through an isothermal core-filament simulation 
H$_2$ number density cube 
\citep{2016arXiv160102018H}
after about 1 Myr of filament 
evolution aiming to provide a simple model for the cores in 
L1517. The simulations were performed
in 3D with the RAMSES code
\citep{2002A&A...385..337T}. The cores form on the
dominant fragmentation wavelength of a gravitational instable, infinite
filament without magnetic fields. The resulting velocity structure of
the core-forming motions is in excellent agreement with the observed 
velocities in L1517.

%-----------------------------------------------------------------------
%-----------------------------------------------------------------------
%-----------------------------------------------------------------------
\begin{figure}
\vbox{
\includegraphics[width=9cm]{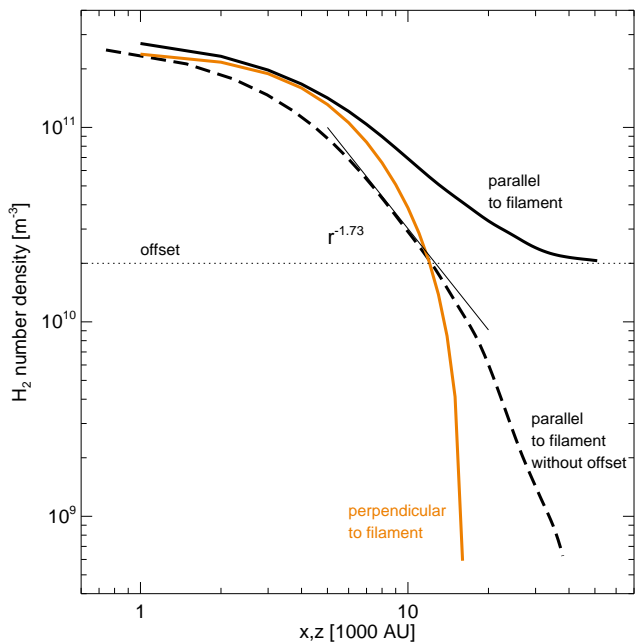}
     }
    \caption{
Cuts through the H$_2$ number density RAMSES results along
the filament axis (thick black), 
with the maximal filament density subtracted (thick, dashed),
and perpendicular to the filament axis (orange). The thin solid line 
indicates the steepness
of a powerlaw with the exponent -1.73 which approximates the dashed line in
a region around 10 kau.
            }
    \label{RAMSEScuts}
\end{figure}
%-----------------------------------------------------------------------
From the inner parts outwards, the core density structure evolves from 
spherical to elliptical symmetry.
The white circle indicates the
radius where the density deviation from spherical symmetry exceeds 50\%.
This radius is in this case about 4/5 of the filament width.
As visible in 
Fig.~\ref{RAMSEScuts},
cuts starting in the core center
perpendicular to the filament axis reach this value at about 
10 kau (orange line). 
The thick black line gives the cut along the filament axis which
reaches the maximal filament density value (thin dotted 
horizontal line) between 15 and 20 kau.
Subtracting this value leads to the contribution of the core 
along the filament (dashed line).
While the perpendicular cut intermixes the spatial gradients of core and 
filament, a horizontal cut is the line with the smallest gradients in 
the filament contribution, and by subtracting the filament maximal value
gives the best measure of the core contribution.
Indeed, subtracting the filament contribution, the simulation data
show a radial power law in the outer core parts where disentangling
core and filament is more difficult (dashed line). The exponent of the power law -1.73
is not far from the simple isothermal sphere value -2 (see thin line).

\subsection{Choice of the cut in the plane-of-sky}
With these findings in mind, we designed the modeling of the core 
located in the filament
based on cuts through the {\it Herschel} maps along the filament axis.
%-----------------------------------------------------------------------
\begin{figure}
\hbox{
\includegraphics[width=8cm]{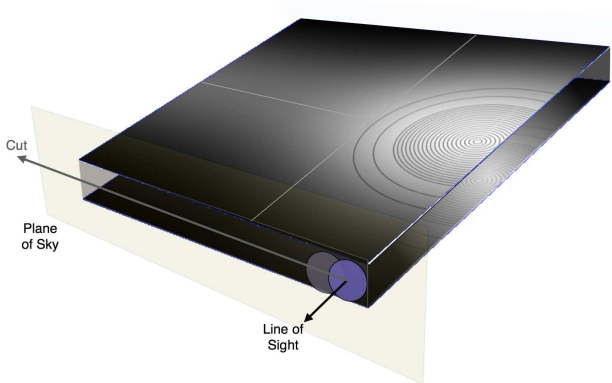}
     }
\caption{
Sketch of the horizontal cut along the filament 
main axis in the plane-of-sky. The cut has a height corresponding to
the beam size at the considered wavelengths. The density in the plane of
lines-of-sight is indicated in grey-scale along with the shells of
the spherical density model.
        }
\label{sketch}
\end{figure}
%-----------------------------------------------------------------------
The basic approach is sketched in Fig.~\ref{sketch}. 
We consider a plane that contains the filament axis and the line-of-sight
and a core that is optically thin at all considered FIR/mm wavelengths.
We assume that the
radiative contribution of the filament along lines-of-sight 
behind and in front of the core can be approximated by the contribution
by the filament outside the core.
To determine the contribution of the core, the filament contribution is
subtracted.

\begin{figure}
\vbox{
\includegraphics[width=9cm]{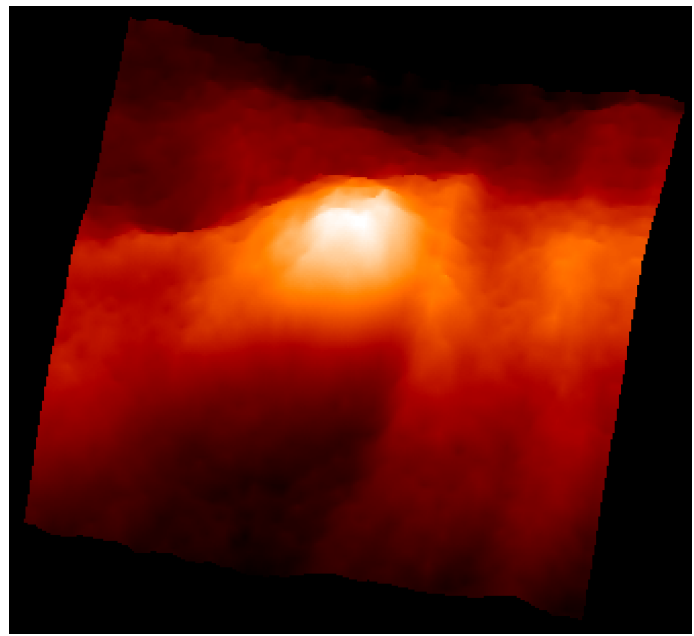}
     }
    \caption{
	    Surface representation of the L1689B {\it Herschel} PACS map at 
     160 $\mu$m amplifying the substructure. 
     The main filament axis is rotated horizontally and the core is
     in the center. Two narrow filaments are visible on the right 
     perpendicular to the main filament, and the filament shows a narrow
     substructure along the main filament axis west of the core. There
     is also sub-structure visible in the central core region at this
     wavelength.
            }
    \label{surface}
\end{figure}

However, 
the core-filament system L1689B shows more complexity than this simple
model. As visible in the surface representation of the map at 
$\lambda$=160 $\mu$m (Fig.~\ref{surface}), the filament has sub-structure
with smaller filaments running perpendicular to the main filament 
to the right of the core. 
Moreover, also the core center shows substructure but since this structure
is not seen at longer wavelengths, we interpret it as a low density warmer sub-filament
behind or in front of the core.
%-----------------------------------------------------------------------
%-----------------------------------------------------------------------
Since there is less confusion from sub-structure to the East of the
core, {we start with modeling the cut} from the center to the West
and discuss other cuts in Sect.~\ref{East}.
%-----------------------------------------------------------------------
\begin{figure*}
\hbox{
\includegraphics[width=6cm]{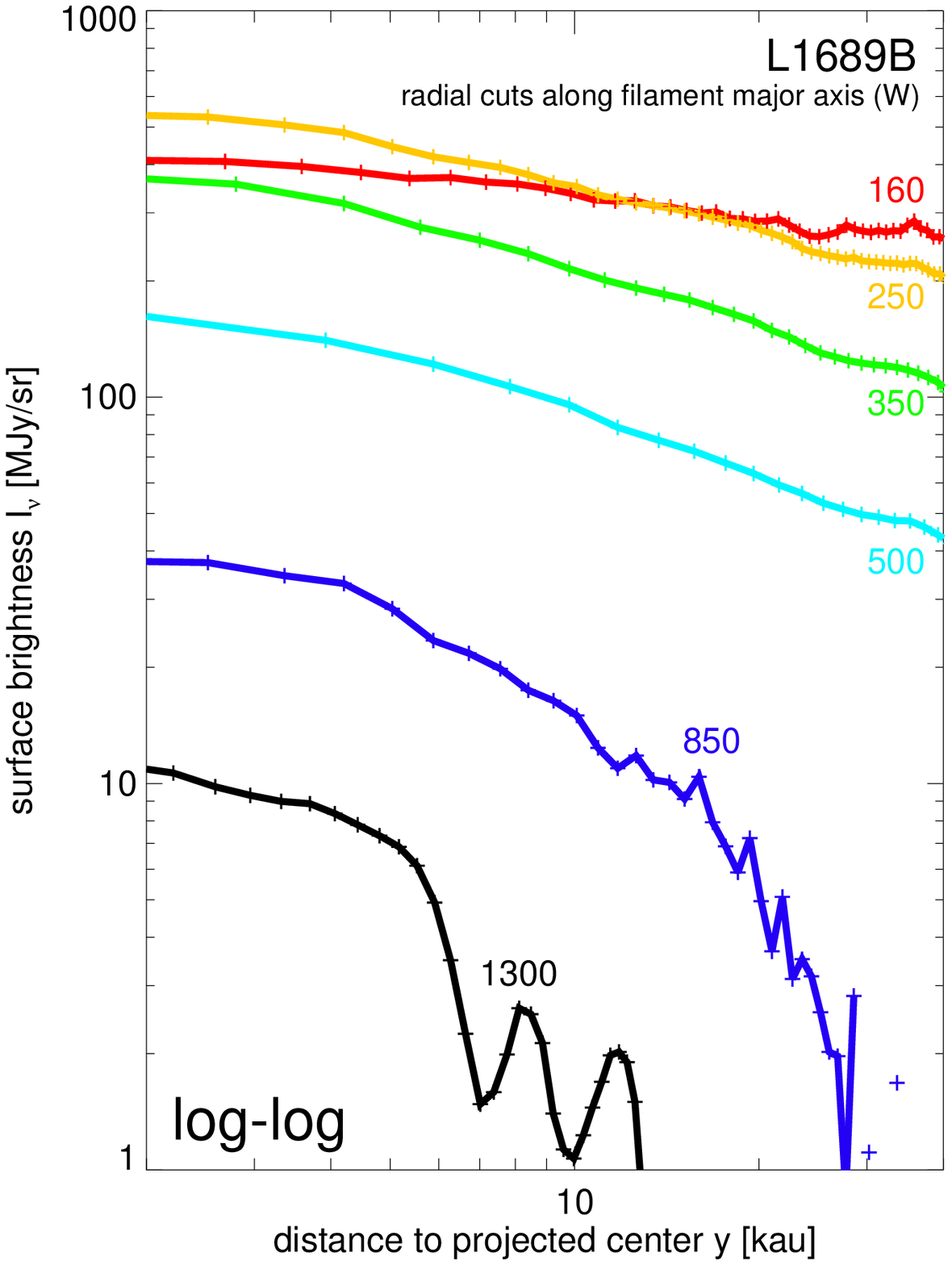}
\includegraphics[width=6cm]{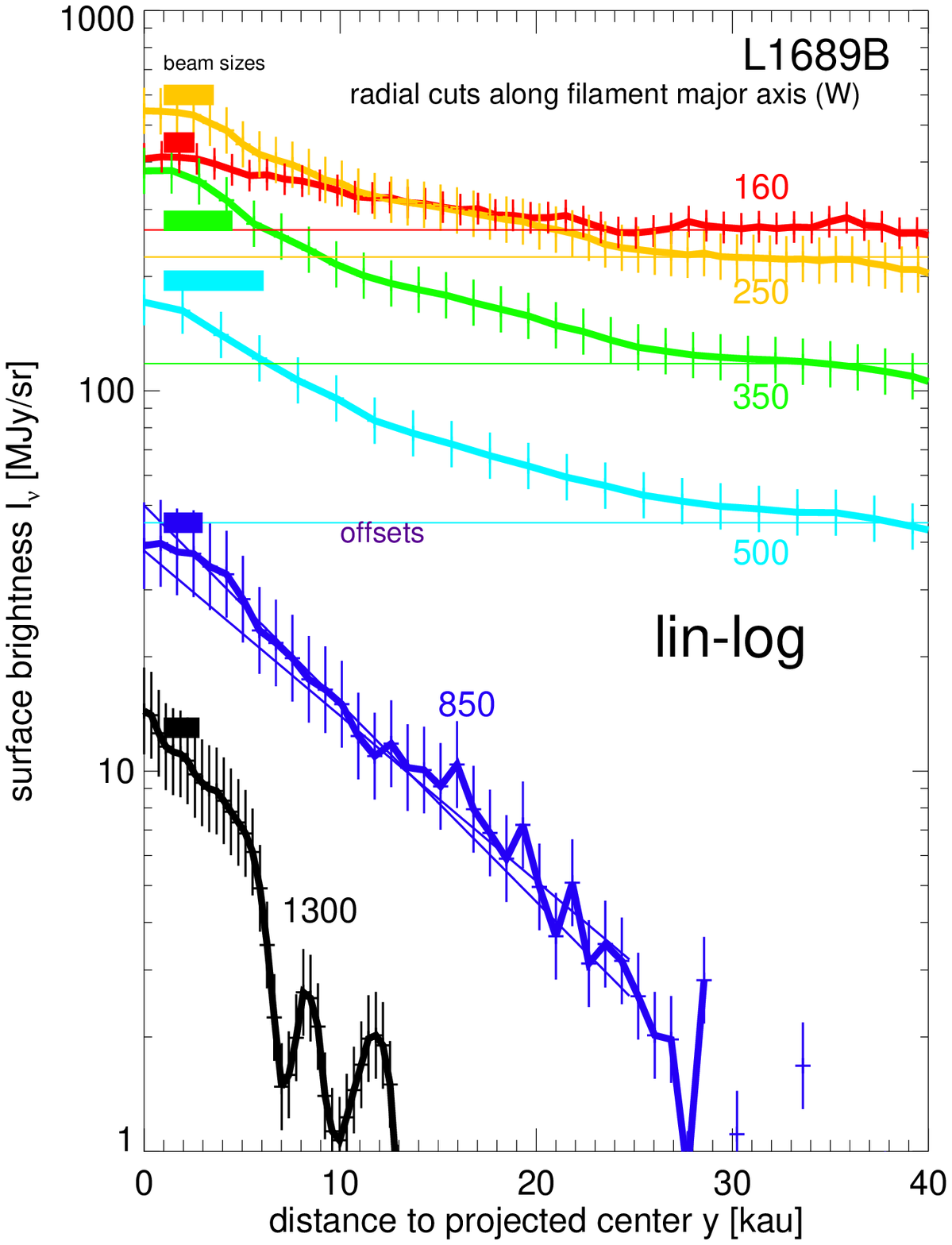}
\includegraphics[width=6cm]{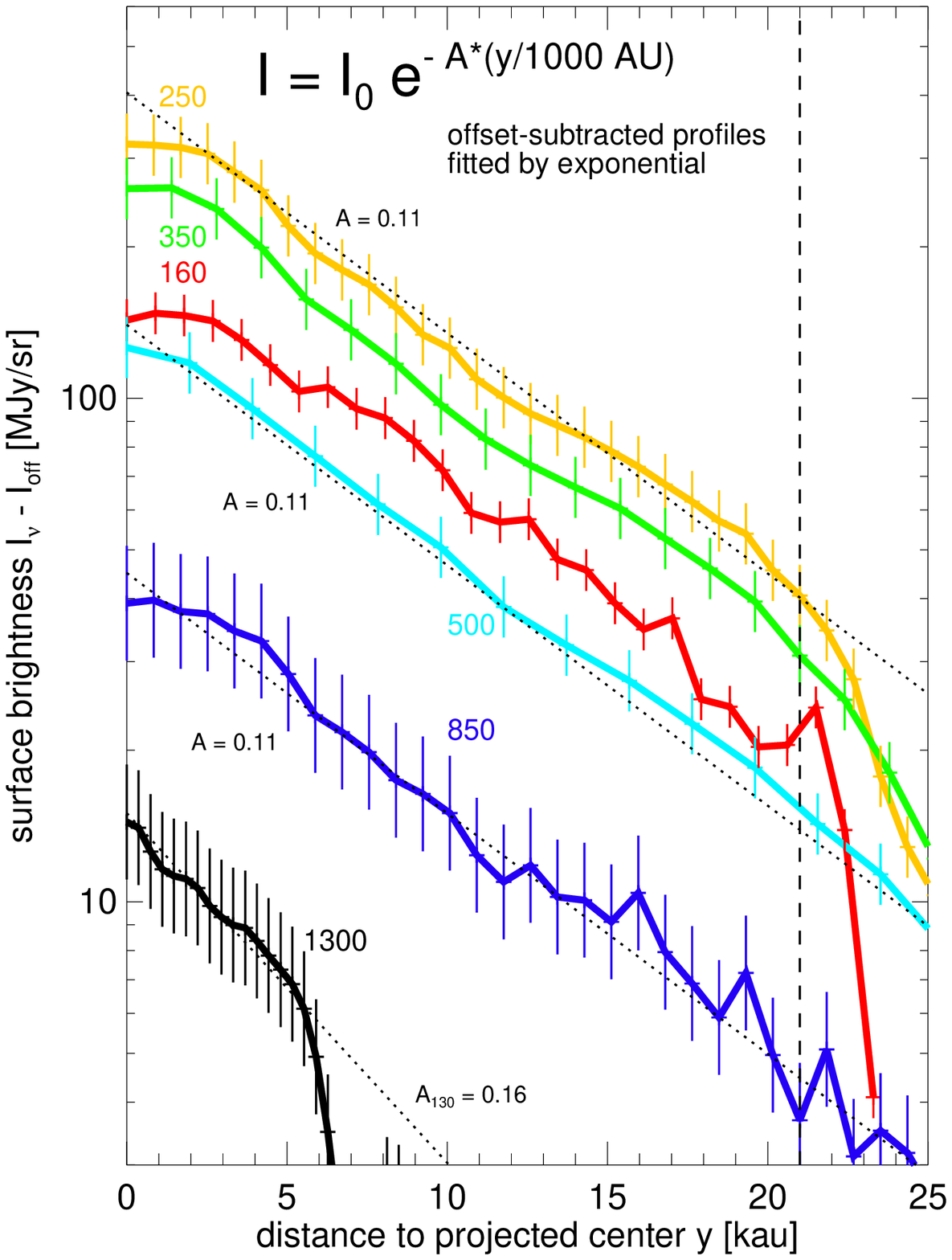}
     }
\caption{
	{\it Left and middle:}
	Radial cuts of the {\it Herschel} maps of L1689B in western direction
	from the core center. The numbers indicate the wavelength. 
	Calibration error bars are added for the middle panel.
        The exponential shape of the 850 $\mu$m profile is better 
	visible in the lin-log representation.
	The horizontal thin lines indicates the offset surface brightness
	from the filament.
	The two thin blue lines estimate the slope variation in the
	800 $\mu$m-profile.
	{\it Right:}
	After subtracting the offsets, all profiles are close to an
	exponential function with approximately the same exponent $A$.
	The dotted lines show the exponential function with $A=0.11$.
	The dashed vertical lines give the distance where deviations 
	from the exponential occur.
        }
\label{three}
\end{figure*}
%-----------------------------------------------------------------------

\subsection{Filament removal unfolding a simple core structure}
The left panel of Fig.~\ref{three} shows the resulting surface brightness profiles
at the various wavelengths as a function of the distance in the western direction from the core center
without any filament subtraction
in log-log-representation.
The {\it Herschel} profiles up to 500 $\mu$m merge into the almost constant 
filament surface brightness at large ($>15$ kau) distances. 
With the declining relative contribution to the surface brightness of the 
filament at longer wavelengths, more insight can come from 
submillimeter images.
Therefore also cuts from the 850 $\mu$m SCUBA-2 and 1.3 mm map are included in our analysis.
The SCUBA-2 data are part of the JCMT Gould Belt survey 
\citep{2007PASP..119..855W}, 
and are described in 
\citet{2015MNRAS.450.1094P}. 
The IRAM 30m data are taken from 
\citet{1996A&A...314..625A}. 
These ground-based maps are not very sensitive to the
the filament's emission, 
as observing techniques to remove atmospheric 
emission also filter out large spatial scales.
Indeed, especially the 850 $\mu$m profiles show an exponential drop 
without a filament contribution
at larger distances. 

A clearer picture emerges when plotting the same data in log-lin-representation
(middle panel of Fig.~\ref{three}).
The 850 $\mu$m profile which should mainly show the radiation from the
core is well-represented by a straight declining line and therefore
can be approximated by an exponential function. 
Following our modeling approach, we approximate the offset in surface brightness 
by the radiation from the filamentary dust behind and in front of the core by 
the surface brightness that we have measured in the filament near but outside 
the core (the approximately flat profile parts 
in the middle panel of Fig.~\ref{three} labeled "offsets").
We also added calibration error bars and two thin blue lines to estimate the slope
range at 850 $\mu$m for the left figure.

The result of the subtraction is shown in the right plot of the same figure.
We note that the data are shown only up to 25 kau in the left panel 
since according to Fig.~\ref{comparable} the contribution from the core
becomes negligible for larger distances.

Interestingly, the surface brightness profiles at all wavelengths follow the same pattern and
can be well fitted by a simple exponential function of the form
\begin{equation}
I(y)=I_0\ \exp\left[-\frac{Ay}{{\rm kau}}\right].
\end{equation}
The exponent shows little variation around $A=0.11$ (dotted lines, with
a variation of 0.1 to 0.12 read, e.g., from
the middle panel for the 800 $\mu$m-profile)
except for the 1.3 mm profile which is slightly steeper $A=0.16$.
As mentioned in \cite{1996A&A...314..625A}, 
the dual-beam observing technique used to map the core at 1.3\,mm induces a filtering 
of the large spatial scales, which leads to an underestimation of the total flux 
as well as a steepening of the radial intensity profile.
Hence, we did not further consider the 1.3\,mm data to avoid introducing large 
errors from a misinterpretation of the surface brightness loss. 
This loss of flux at large spatial scales also affects the SCUBA-2 data but 
to a much smaller extent, since spatial scales up to 5$\arcmin$ can be recovered 
\citep{2015MNRAS.450.1094P}, which is larger than the size of the core.
The vertical dashed line indicates where deviation from the exponential shape
occur.

%-----------------------------------------------------------------------
\begin{figure}
\vbox{
\includegraphics[width=9cm]{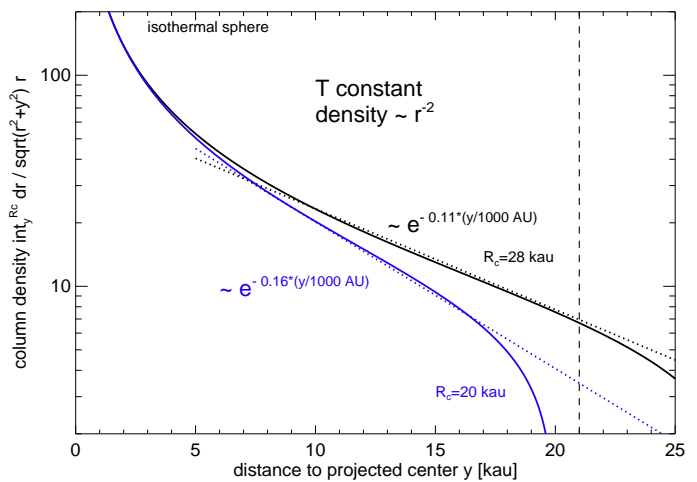}
     }
\caption{
The line-of-sight integral through an iso-thermal sphere (solid) can
be approximated in large parts by an exponential (dotted).
        }
\label{expfit}
\end{figure}
%-----------------------------------------------------------------------
Exponential profiles are not unknown when it comes to integrating
spherically symmetric density structures in order to obtain a column density profile.
In Fig.~\ref{expfit} we show the column density profile of an isothermal
sphere with a power law density profile $\propto r^{-2}$, and truncated at radius $R_c$
(solid lines). 
Between the steep rise caused by the singularity at $r=0$ and the
cutoff at the core outer radius $R_c$ the column density profile is
well-described by exactly the same exponential shape that we observe
in the L1689B surface brightness profiles (dotted lines).
For more general density profiles with a power law of the form
$r^{-\alpha}$,
the factor $A$ is numerically found to depend on the ratio
$\alpha/R_c$. 
For $\alpha=2$, $A$ reaches 0.11 as observed for outer core radii of
about $R_c=28$ kau.
We note that at this value the column density vanishes, but at shorter
distances of about 21 kau for $A=0.11$ deviations from the exponential
become visible (dashed line). We therefore restrict modeling of the core up to 
21 kau.

In the more realistic model of a Bonnor-Ebert sphere with a central flattening, the pole
at $y=0$ disappears and depending on the radial extent of the flattened region the
agreement with the exponential function can extend to smaller distances.

We conclude that for radii > 5 kau, 
the core shows approximately the same surface brightness distribution
at all wavelengths once the filament contribution is removed following
the simple scheme that we have described.
The resulting surface brightness profiles already enable a direct interpretation
without any radiative transfer modeling. They are consistent with radiation
from an isothermal spherical core with a radial density profile following an
$r^{-2}$-dependence. 
calculations.
In the following we use the exponential approximation for the profiles
with an outer radius of $R_c=28$ kau between 5 kau and 21 kau.
As visible from Fig.~\ref{three}, the deviations of the observed surface
brightness values from the exponential approximations are of the order
of the calibration errors which enter the radiative transfer
modeling performed in Sect.~\ref{Deri}. We therefore consider that the
impact of this approximation on the derived densities and temperatures
is comparable to that caused by the calibration errors.

\subsection{Modeling for other cuts than the Western profile}\label{East}
\begin{figure}
\includegraphics[width=9cm]{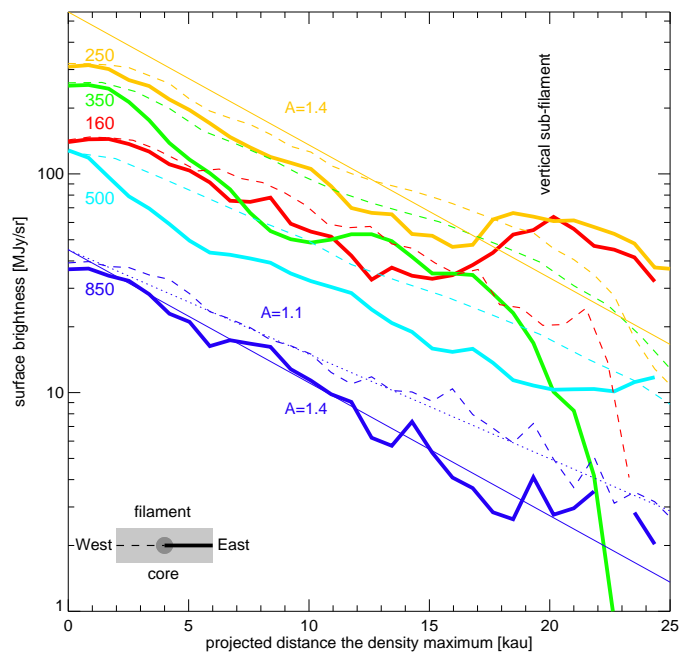}
\caption{
	Offset-subtracted surface brightness
profiles for the Eastern cut along the filament axis
(thick) in comparison to the Western cut (thin dashed, Fig.~\ref{three}).
The exponential fit leads to steeper gradients ($A=1.4$) than for the
Western cut. The surface brightnesses at 160 and 250 $\mu$m
are enhanced due to a vertical filament in the region around 20 kau
distance.
        }
\label{eastcut}
\end{figure}

The proposed method based on cuts along the filament axis can also
be applied to the other direction (East).
Fig.~\ref{eastcut} shows the resulting offset-subtracted surface
brightness profiles for a cut in Eastern direction (thick line)
for the various maps. 
For the sake of comparison we have added the Western cuts as thin 
dashed line. Again, the profiles reveal the same exponential shape
for all wavelengths but with a slightly steeper gradient $A=1.4$ than
for the Western cut. 
It is more difficult to read the distance at which deviations from
the exponential approximation occur due to the vertical filament that affects the
map at distances of about 20 kau (visible in the 160 and 250 $\mu$m
profiles). The only profile showing a clear decline is the 350 $\mu$m
profile starting at about 19 kau. 
According to Fig.~\ref{expfit} this would correspond to a density
gradient of about -2 again as for the Western cut. But due to the
confusion with the sub-filament this conclusion is to be handled with
caution.

\begin{figure}
\includegraphics[width=9cm]{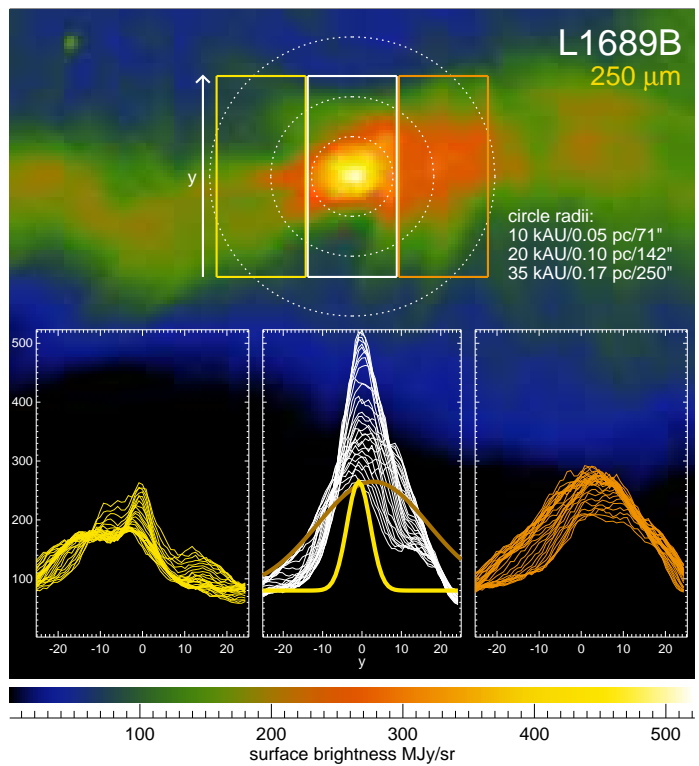}
\caption{
	Vertical surface brightness profiles (direction y) in the 250 $\mu$m
map in three boxes left, central, and right around the core.
The left box shows the narrow filament maximum (indicated again in
the middle box as yellow line). The right box shows a broader maximum
(at about the same level, see brown line in the middle box).
The central cuts are shown as white lines in the middle box.
Dotted circles indicate the distance from the center.
        }
\label{fig0}
\end{figure}

It could also be considered to model a vertical cut. In a similar manner
as the horizontal cuts, the filament off-core could be subtracted from
a vertical central cut through the core to retrieve the core contribution.
For a symmetric core filament system like the simple density structure
seen in the RAMSES simulations (Fig.~\ref{RAMSESsym}), this would lead
to an approximate estimate of the core density.
The L1689B filament, however, shows stronger asymmetries. Indeed, the 
filament profiles on both sides of the cores differ substantially as
visible in Fig.~\ref{fig0}. Vertical cuts through the 250 $\mu$m map
divided into a Western, a central, and an Eastern box show a narrow
maximum in the surface brightness to the West while the Eastern profiles
are broader. In the central inlet, the white vertical profiles are compared to 
approximate fits of the left and right central profiles.
In this case it is difficult to decide which profile would have to be
subtracted to arrive at the core surface brightness profile.

Therefore, we rely in our analysis on the Western surface profiles
noting that the Eastern profiles appear to have a similar underlying
density structure.
%-----------------------------------------------------------------------
%-----------------------------------------------------------------------
%-----------------------------------------------------------------------
%-----------------------------------------------------------------------

%+++++++++++++++++++++++++++++++++++++++++++++++++++++++++++++++++++++++
\section{Basic numerical method: peeling radiative transfer}\label{Nume}
\begin{figure}
\vbox{
\includegraphics[width=9cm]{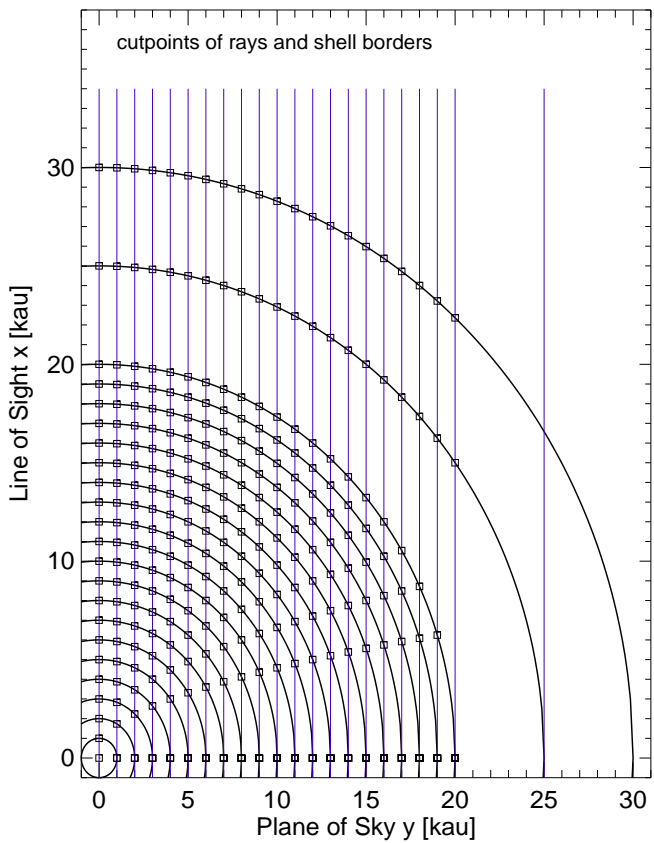}
     }
\caption{
	Illustration of the used shell and ray structure including
	the cutpoints. Moving inward more and more shell contributions
	enter the observed surface brightness.
        }
\label{cuts}
\end{figure}

\begin{figure}
\vbox{
\includegraphics[width=9cm]{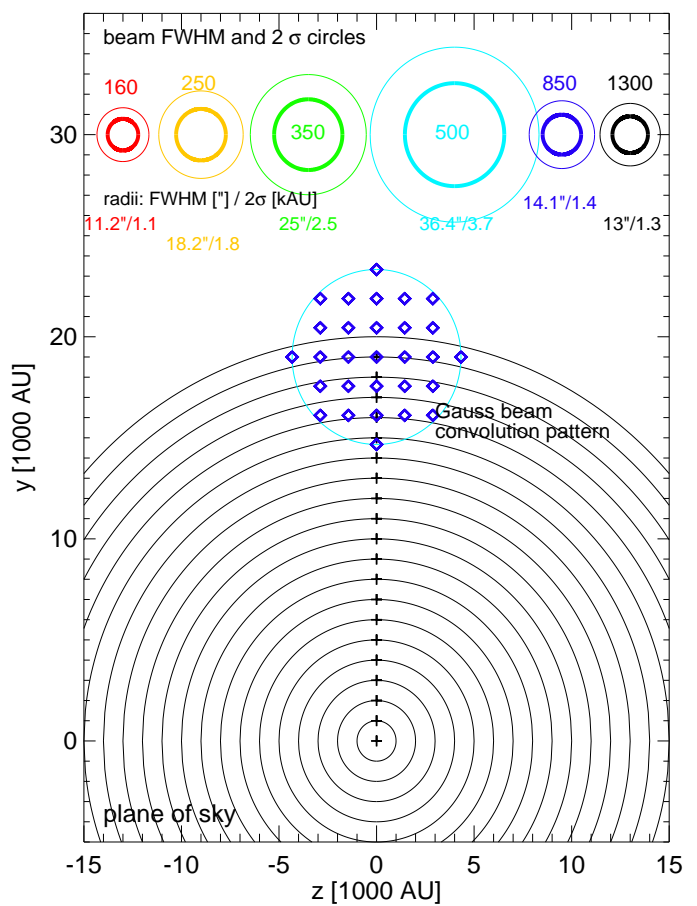}
     }
\caption{
	Comparison of the shell structure and the telescope beams at 
	the different wavelengths. The size of the beam 
	(approximated by a Gaussian with the variance $\sigma^2$)
	is illustrated
	both by the full width half maximum (FWHM) and $2\sigma$ circles.
	Gaussian convolutions are performed using a 29 point stencil
	shown in the figure center.
        }
\label{beams}
\end{figure}

Our calculations of the core density and temperature are adapted to 
the centrally-condensed structure of L1689B. 
The RT calculations are simple since L1689B is optically
thin for the observed wavelengths discussed here ($\lambda > 160\ \mu$m) 
which leads to a decoupling of 
the radiative contributions of all parts in the core-filament
system (as already pointed out by R+14). 
For line transfer in core-filament systems we also refer to
\citet{2012ApJ...750...64S} and for continuum transfer in filaments to 
Malinen 2012 \citet{2012A&A...544A..50M}.
We use forward ray-tracing where the term {\em forward} 
is referring to first choosing a density
and temperature and then determining the observed radiation. 
In this work, we only use a set of spherically symmetric shells
but the 
used ray-tracing method described here 
works with any sort of centrally condensed layer 
structures and the cutpoints and segment lengths of 
the rays and layers are easy to determine even for complex layer structures. 
For the plane sketched in Fig.~\ref{sketch} the shells and rays are shown in
Fig.~\ref{cuts}. For simplicity we show rays for a cut in the Eastern direction
while the cut through L1689B runs westward.
We chose to describe the core with 18 concentric shells of 1 kau thickness,
two outer shells of 5 kau thickness, all with constant density and temperature. (Fig. 8).
The contribution to the emission of each shell at each radius can be easily calculated, 
starting from the outermost shell inwards ("peeling"). 
The density and temperature of shell $i$ can be determined by single 
temperature SED fitting of the shell contribution to the overall intensity at 
radius $i-1$.
%-----------------------------------------------------------------------
%-----------------------------------------------------------------------
We investigated the impact of beam convolution comparing the un-convolved
profiles with beam-convolved cuts with a 29 point stencil Gaussian
beam pattern (Fig.~\ref{beams}) out to 
2 $\sigma$ of the beam for each wavelengths. Like R+14
we find no significant effect for the outer parts of the core
and the beginning filament. 
Contrary to their work, we however find a significant smearing in the
central parts and therefore use beam convolution for the inner five shells.

%-----------------------------------------------------------------------
\begin{figure}
\hbox{
\includegraphics[width=4.5cm]{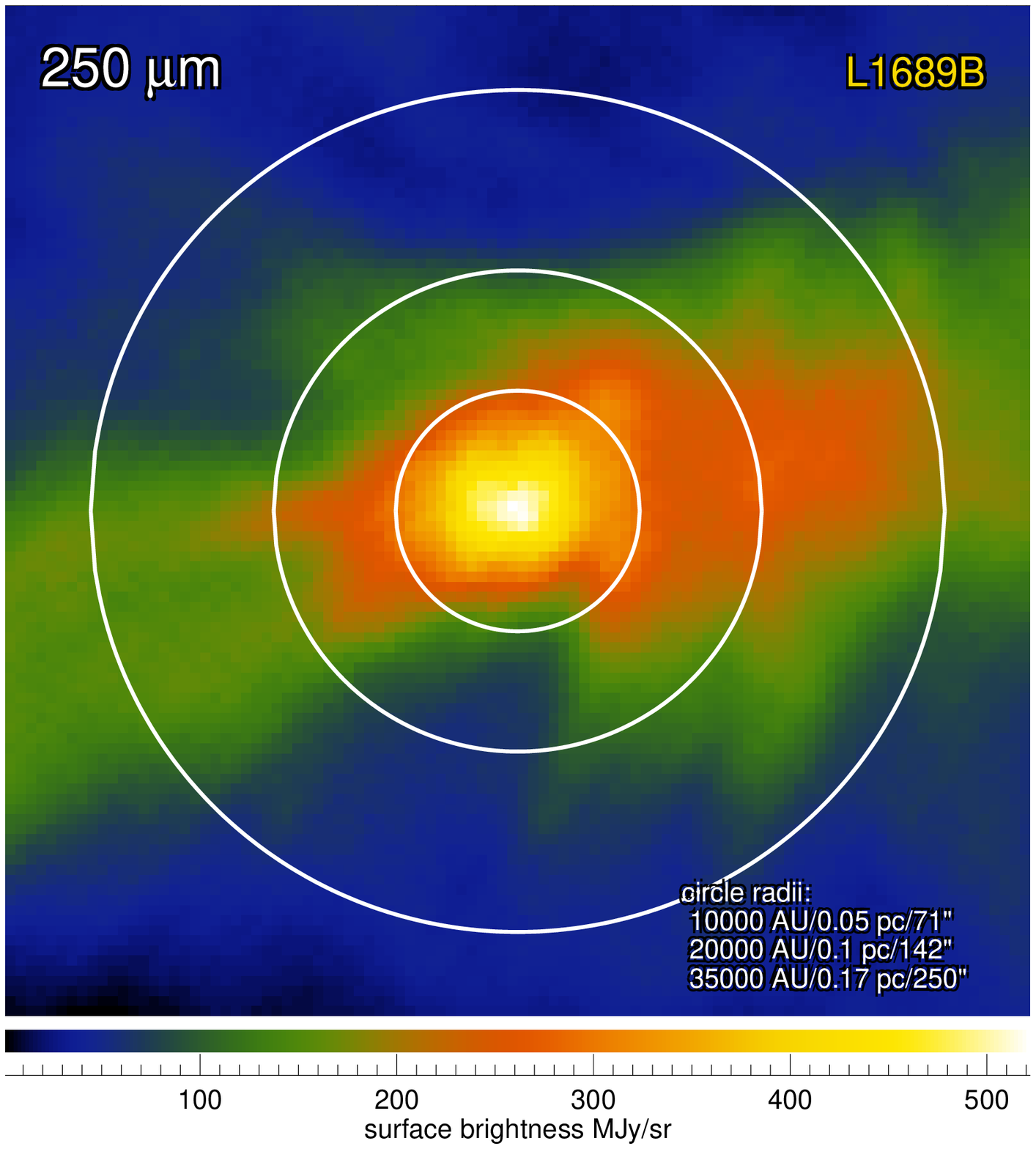}
\includegraphics[width=4.5cm]{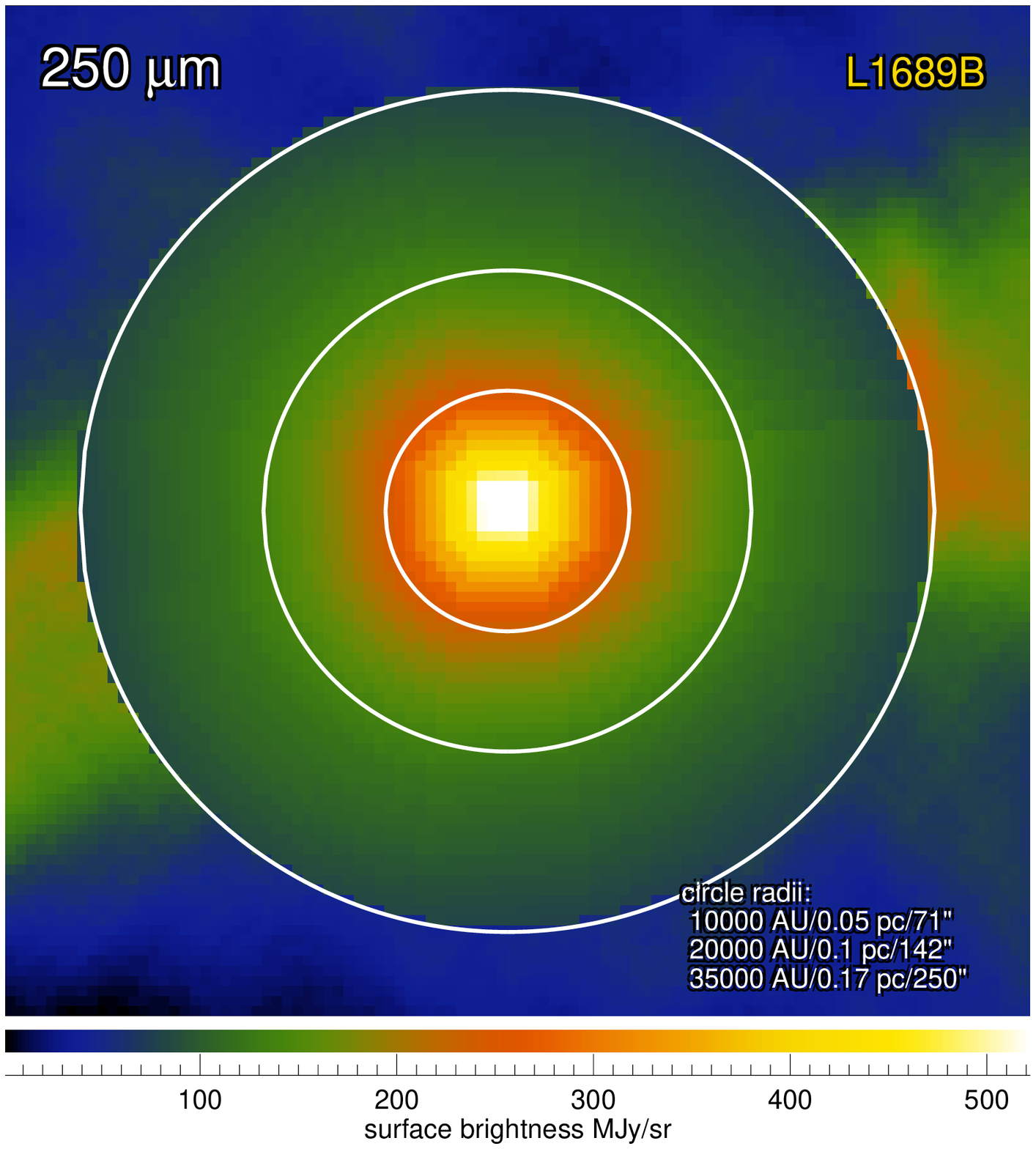}
     }
		     \caption{
			     The figure depicts the {\it Herschel} 250 $\mu$m map of L1689B (left) and the strong
alteration by an azimuthal average as performed in the R+14 model (right),
mixing filament, core, and filament-external contributions. White circles
indicate various radial distances from the center.
   }
     \label{leftright}
   \end{figure}

%-----------------------------------------------------------------------
This approach differs from the inverse RT
used in
R+14 
based on an Abel
transform inverting
the problem directly providing the unknown density and temperature. 
This method can also be generalized 
to elliptical shells. 
Particular differences in the modeling are:
\begin{itemize}
\item
Forward ray-tracing does not require to smear all maps to the same
(coarse) resolution of the 500 $\mu$m maps which causes a loss of spatial information
especially in the central parts.
Instead individual beam smearing is performed after the theoretical radiation
field has been determined from the chosen temperature and density grid.
\item
R+14
derive their density and temperature profiles from
azimuthally averaged surface brightness, and use the deviations from the average along different directions
to estimate uncertainties in their profiles. 
With this approach, it is not clear how to define
a radius where the core ends and where the model is 
averaging filament matter with non-related low density matter outside the core-filament system.
The substantial alteration of the observed surface brightness by
the azimuthal average is illustrated in Fig.~\ref{leftright}. 
The observed 250 $\mu$m map is compared to the same map with direction-averaging 
out to 35 kau as in R+14.
In contrast, our core-filament border estimates in Sect.~\ref{borders} point
to 9 to 15 kau as the outer core radius.
\item
The central part of the core is the most interesting region in terms
of the basic unanswered questions about the onset of star formation. 
It also hosts the coldest dust in the system which by the simple
formalism of black body radiation is hardest to see (a drop of 1 K decreases
the maximal Planck radiation by a factor of 2 due to its 
$T^5$-dependence). The uncertainty in the density and temperature arising from this 
lack of sensitivity is described in the next section.
\end{itemize}
%-----------------------------------------------------------------------
%-----------------------------------------------------------------------

%+++++++++++++++++++++++++++++++++++++++++++++++++++++++++++++++++++++++
\section{Deriving the density and temperature structure along the western cut}\label{Deri}
\subsection{Grid calculations for the inner shells}\label{Grid}
There are a few circumstances that make the detection of cold dust in
the core center more difficult. Lines of sight through the center always see emission
from all layers. In competition with the outer warm (> 10 K) and thus
brighter dust, the colder central dust must compensate with a higher density to become
observable. Moreover, 
because of the limited angular resolution of smaller telescopes like {\it Herschel}, 
the flux coming from warm and cold regions is spatially averaged, 
and gradients are smoothed out, especially at longer wavelengths like 500\,$\mu$m.
Consequently, ground-based observations at longer wavelengths with larger telescopes that
have a smaller beam are better probing the central region. 

To determine the density and temperature structure along the Western cut of the core, we have treated
separately the inner five shells (spanning 5 kau in radius) and the outer shells beyond 5 kau.
For the outer region, we used the exponential approximations of the
observed surface brightness as shown in Fig.~\ref{three} down to 6 kau, and applied the method described in Sect.\,3.
For the inner shells, for which convolution with the telescope beam has to be taken into account we have run forward
radiative transfer over a full grid of densities and temperatures
with 10 values in each of the 10 parameters (5 densities and 5 temperatures), 
deriving in each case the intensity after convolution with the appropriate beam.
Since the precise shape of the error distributions of the observed
surface brightnesses is not known, we have taken the simple approach to
accepted all solutions fitting all
surface brightness profiles simultaneously within the calibration errors
of the bands estimated to be 10\% for SPIRE, 15\% for PACS, and 30\% for SCUBA-2, see 
\citet{2016A&A...587A..26B}).
For the sake of comparison, we have the same dust opacities as R+14
(HGBS opacities, and a gas-to-dust ratio 100).

\subsection{Grid calculations for the outer interstellar radiation field}\label{GridISRF}
For the inner shells, we found a range of density and temperature profiles
consistent with the data. Indeed, increasing the central density (and hence obtaining a steeper profile) can be compensated by
lowering the temperature. 
However, solutions for the temperature profile should be consistent with a temperature profile obtained by 
calculating the propagation of interstellar radiation through the corresponding density structure.
Hence, we have designed a new synergetic RT approach. 
Since we determined the outer (> 5 kau) density and temperature structure,
we can use this information to estimate the outer radiation field.
The ISRF was parameterized following the
approach by 
\citet{2001A&A...376..650Z}
based on the work by
\citet{1994ASPC...58..355B}.
%-----------------------------------------------------------------------
\begin{figure}
\vbox{
\includegraphics[width=9cm]{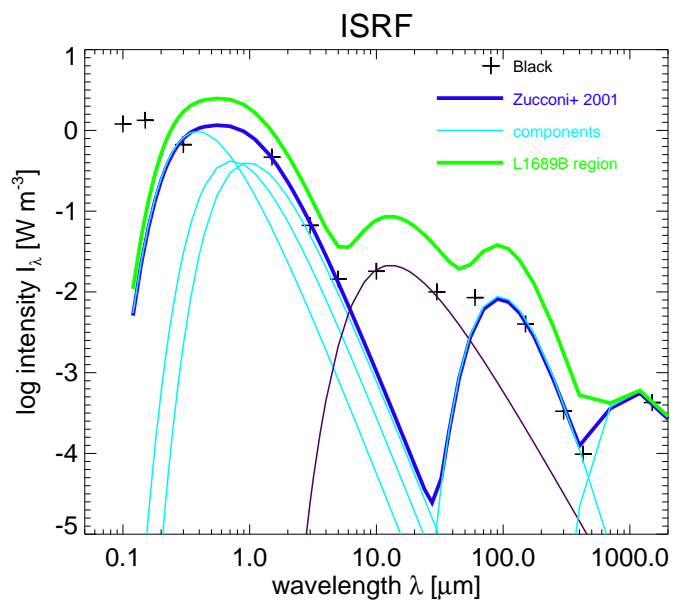}
     }
\caption{
Standard interstellar radiation field for a low-mass
star formation region (blue) after \citet{2001A&A...376..650Z} with its
different components as magenta lines. For comparison, points from the
\citet{1994ASPC...58..355B}
field are shown as plus signs.
A mid-infrared component was added (thin black line) to arrive at these values.
The green line gives the local ISRF of L1689B.
        }
\label{ISRF}
\end{figure}
%-----------------------------------------------------------------------
Fig.~\ref{ISRF} presents their ISRF spectrum (blue)
composed of the individual components (magenta)  
and the \citet{1994ASPC...58..355B} data (black crosses).
As pointed out by \citet{2001A&A...376..650Z}, their ISRF is a minimal field not containing
the likely addition of a mid-infrared (MIR) component due to warm dust in
star formation regions. We therefore added this component (thin black)
to reproduce
the 
\citet{1994ASPC...58..355B} data. Based on this ISRF we have run a large grid
of RT calculations determining a new temperature
profile using the density structure derived for L1689B by "peeling" RT, as described in Sect.\,3.
We used the standard dust opacities by 
\citet{1984ApJ...285...89D} which we merged into the HGBS opacities
in the FIR and submillimeter for consistency.
In each run we have varied each of the components of the ISRF by factors ranging from 0.1 to 10:
stellar (first three components),
warm dust (peaking around 10 $\mu$m), and cold dust (peaking around 100 $\mu$m).
The spread in possible ISRFs fitting both the peeling RT temperature profile
and the observed surface brightness profiles was narrow. The ISRF near L1689B is best 
described by an ISRF with twice the stellar component and four times the 
warm and cold dust components (green).

Based on this local ISRF for L1689B, we have re-calculated also the temperature
of the inner five shells for each of the
formerly derived density profiles. 
By accepting only solutions where the two differently derived temperatures
match, we were able to extract the solution subspace that contains 
density and temperature profiles both fitting all surface brightness
profiles and being consistent with a single outer ISRF.

%-----------------------------------------------------------------------
\begin{figure}
\vbox{
\includegraphics[width=9cm]{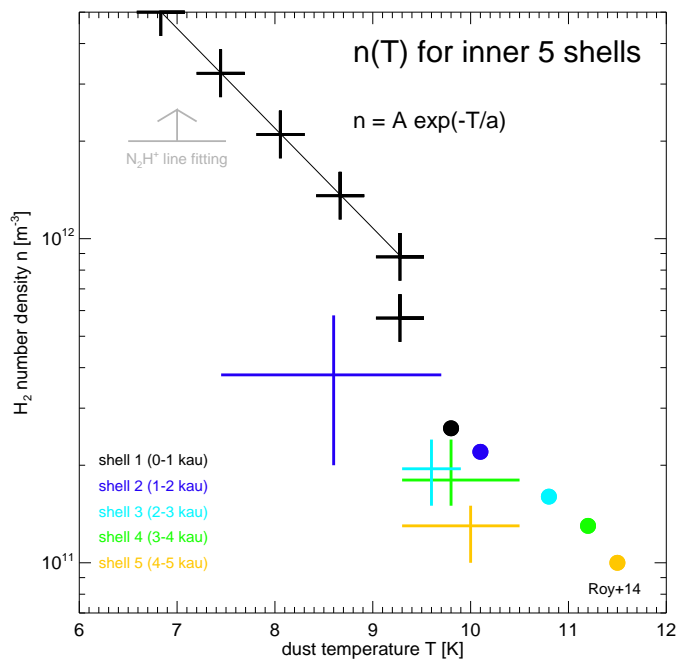}
     }
     \caption{
     H$_2$ number density- dust temperature for the five inner shells
     of L1689B with a size of 1 kau each. The crosses mark the ranges 
     for which all profiles can be fitted within the calibration errors
     simultaneously. The dots represent the mean values from R+14.
     For the inner shell, $n$-$T$ pairs are possible along an exponential
     function (A=6.4$\times$10$^{14}$ m$^{-3}$ and a=1.4 K).
     The grey lower limit indicates densities which are compatible with
     line emission modeling as mentioned in Sect.~\ref{Disc}.
             }
\label{nT}
\end{figure}
%-----------------------------------------------------------------------

Fig.~\ref{nT} shows the corresponding density-temperature solution space 
(hereafter n-T) for each of the five inner
shells (color-coded) as crosses. 
In the two inner shells (blue and black) some ambiguity remains:
in the second shell the temperature (density) ranges between 7.5 and 9.6 K 
(2$\times$10$^{11}$ m$^{-3}$ and 5.8$\times$10$^{11}$ m$^{-3}$), respectively - 
lower in temperature and higher in density than values from the mean R+14 (dots).
We note that the mean R+14 values are an average while they also performed fits along
individual cuts. But since these individual profiles were only used
in the error bars we rely on the mean values here.
In the inner shell,  any pair of $n$ and $T$
values obeying the relation $n=A\exp{(T/a)}$ (A=6.4$\times$10$^{14}$ m$^{-3}$ and a=1.4 K) - is still be consistent with
the data and the ISRF.
Our solution includes densities above 10$^{12}$ m$^{-3}$ 
for temperatures < 9 K.
Physically, the ambiguity reflects the fact that with all the warmer shells
contributing to the central line-of-sight, the additional cold dust emission 
contribution
can be kept constant when a higher density is compensated with lower temperature.
The ambiguity is increased when beam-convolution is performed or the maps are
smoothed to the coarse spatial resolution of the 500 $\mu$m {\it Herschel} observations.

\begin{figure}
\vbox{
\includegraphics[width=9cm]{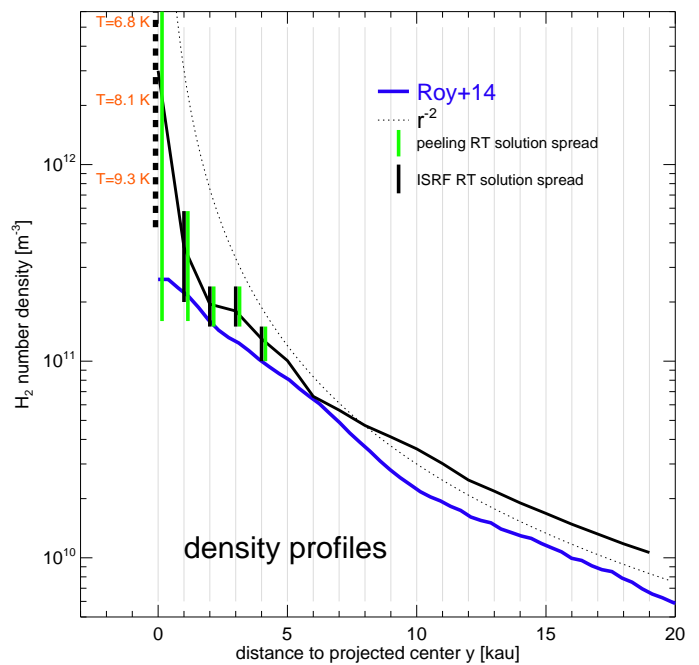}
     }
\caption{
	H$_2$ number density profile for L1689B (black solid).
	Like the profile from R+14 (blue) the outer core parts
	approximately follow an $r^{-2}$ law but arrive at higher
	central densities. The remaining ambiguity in the inner cell
	is shown by green bars. The $n$-$T$-relation of possible 
	inner shell solutions is marked with a dashed bar giving the
	corresponding temperatures.
      }
\label{nprof}
\end{figure}

The resulting H$_2$ number density profile is shown in 
in Fig.~\ref{nprof} (black solid) along with the R+14 profile (blue), 
and an $r^{-2}$-profile (dotted). For the inner cells, the density values fitting
all surface brightness profiles within the calibration errors 
are indicated as green bars. With the constraints derived from the ISRF, the ranges
shrink to the black bars. For the inner cell, the bar is plotted dashed
to indicate that each density within the range is correlated with 
the temperature following the exponential relation visible in Fig.~{nT} for high densities.

\begin{figure}
\vbox{
\includegraphics[width=9cm]{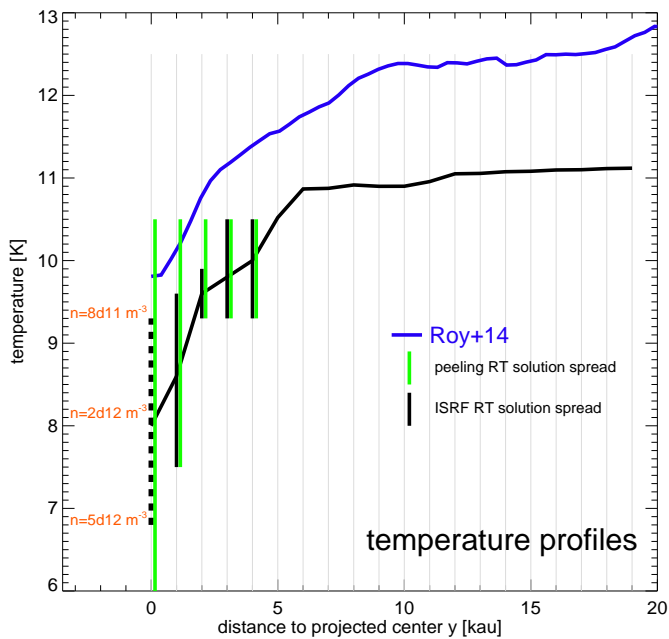}
     }
\caption{
	Dust temperature profile for L1689B (black solid).
	The temperatures are lower than those of R+14.
	The notation is the same as in Fig.~\ref{nprof}.
      }
\label{Tprof}
\end{figure}

With a similar color code, we show the dust temperature profile in Fig.~\ref{Tprof}.
The temperatures are lower than the R+14 mean values, and can
decrease to about 6.8 K in the inner cell.

%-----------------------------------------------------------------------
%-----------------------------------------------------------------------
%-----------------------------------------------------------------------
%-----------------------------------------------------------------------

\section{Discussion}\label{Disc}
In this section, we discuss a number of implications based on our new
density and temperature profiles in compact subsections.
\subsection{Strength of the interstellar radiation field: the standard ISRF amplified
by a factor of 10 is too high}
%-----------------------------------------------------------------------
\begin{figure}
\vbox{
\includegraphics[width=9cm]{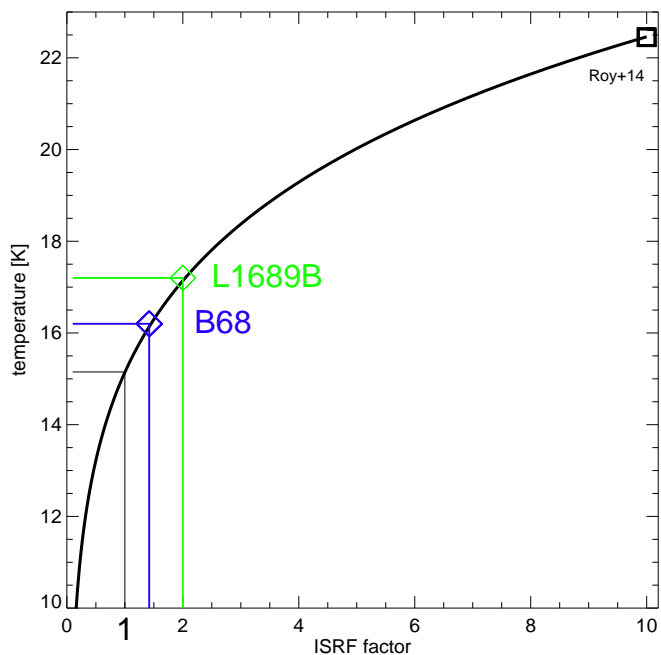}
     }
\caption{
	Temperature of dust fully exposed to the radiation field as
	a function the amplification factor of the standard interstellar
	radiation field. 
      }
\label{ISRFfac}
\end{figure}
%-----------------------------------------------------------------------
In view of the obtained relatively high central temperature of L1689B of 9.8$\pm$0.5 K
R+14 argued that the radiation field near the core may be a factor of 10
stronger than the standard ISRF due to the presence of early-type
stars in the vicinity of the main (L1688) Ophiuchus cloud as treated in \citet{1999A&A...344..342L}. 
This factor might however not apply to the L1689 cloud, which is located further
away from these stars. A simple calculation can be made for the temperature of the dust that is 
on the outskirts of the filament and therefore well-exposed
to the complete radiation field.

Fig.~\ref{ISRFfac} presents the temperature of the dust which
is fully exposed to a scaled-up standard ISRF as a function of the scaling factor. 
The outer temperature near L1689B and B68 can easily be derived from the 
maps obtained by pixel-by-pixel single temperature SED fitting in R+14 
or in Sect.~\ref{pix}. 
This enables us to determine the ISRF scaling factors to account 
for the outer temperatures in B68 and L1689B, 1.5 and 2, respectively.
The value of 2 shown in the figure for L1689B is consistent with the
more precisely determined ISRF in Sect.\ref{GridISRF} of the present work with factors of 2 (stellar), 4 (warm
dust), and 4 (cold dust) since the outer dust is mainly heated by the
stellar component. ISRFs ten times higher would heat
the dust to about 22.5 K which is not observed near L1689B.

\subsection{Mass and central column density: a denser center}
%-----------------------------------------------------------------------
\begin{figure}
\vbox{
\includegraphics[width=9cm]{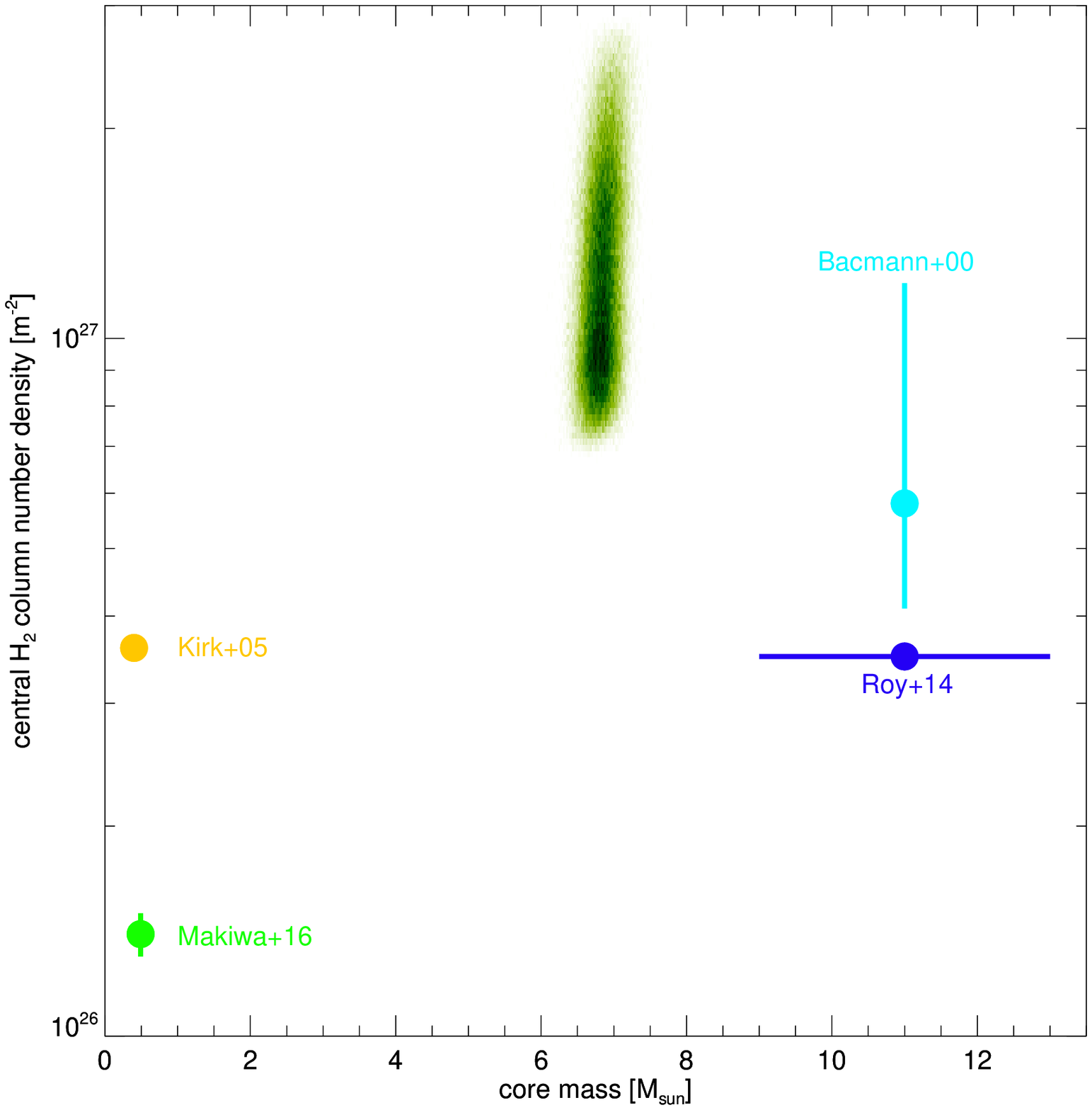}
     }
\caption{
Correlation between the central H$_2$ column number density and the core mass of L1689B. 
Aside of various ranges from the literature, the solution space found
in this work is shown color-coding the number of solutions at about
7 M$_\odot$. The distribution has a longer extent in column density
since it is more affected by the dense inner shells than the mass.
      }
\label{Nm}
\end{figure}
%-----------------------------------------------------------------------
The variation in derived column densities and especially in masses
as mentioned in the introduction is shown in Fig.~\ref{Nm} along with
the our modeling result (both fitting all profiles and consistent with
a single ISRF). We have color-coded the number of solutions which all
are located near 7 M$_\odot$ in mass. As the inner dense shells impact the
column density much more than the mass, the spread in possible column
densities is larger and extend well above the values discussed so far.
The R+14 core mass is too high as it includes filament contributions.
We note that we have used the same dust opacities and source distance
to exclude potential differences from these assumptions. 
A similar argument applies for the mass estimate by 
\citet{2000A&A...361..555B} based on an outer radius of 28 kau.
Their opacities are near the HGBS values used in our work,
but they assumed a distance of 160 pc instead of 140 pc
which increase the mass by about 30\%.
\cite{2005MNRAS.360.1506K}
and
\citet{2016MNRAS.458.2150M}
used slightly smaller distances (130 pc and 120 pc, respectively)
but again comparable opacities, and nevertheless derived much smaller
masses.
Their estimates likely suffer from the single-fit approximation
as shown in the next subsection.

\subsection{Pixel-by-pixel single temperature SED fits do not work
for dense cores}\label{pix}
%-----------------------------------------------------------------------
\begin{figure}
\vbox{
\includegraphics[width=9cm]{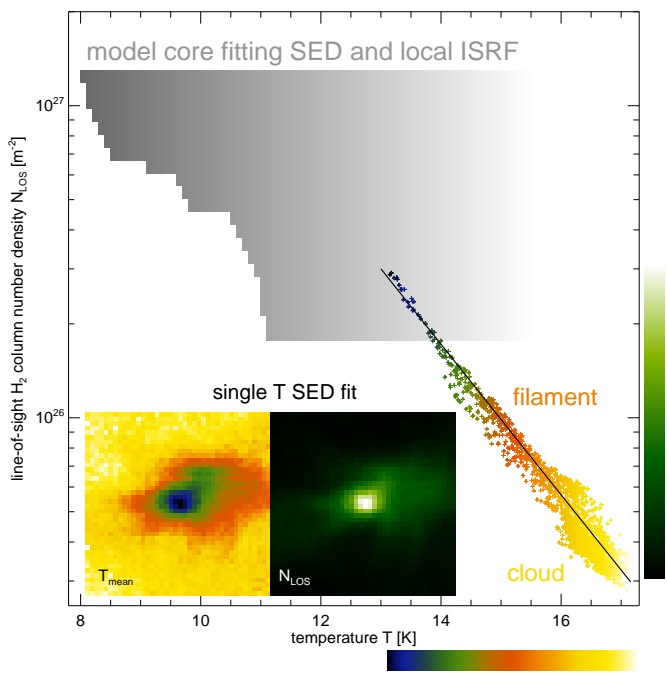}
     }
    \caption{
     Line-of-sight H$_2$ column number density and temperature for 
     the pixel-by-pixel single temperature SED fit (points along
     the line, temperature and column density map shown as inlet).
     The grey-scaled region shows the area where all model
     column densities and temperature solutions are located 
     (the color-coding is based on the logarithmic density at the
     various temperatures entering the column density)
     }
   \label{NT}
  \end{figure}
%-----------------------------------------------------------------------
  Especially in the first {\it Herschel} data releases, pixel-by-pixel single
temperature SED fits of regions with cores have been presented
\citep[e.g.,][]{2010A&A...518L.106K}.
They have turned out to be helpful in characterizing the core environment
where the single temperature assumption is valid, because of the low densities.
Fig.~\ref{NT} shows the temperature and column density maps based on 
single-temperature SED fitting of the
L1689B continuum data as inlet. Clearly, the two maps are correlated in
structure, and indeed the column density-temperature relation shown
is surprisingly narrow (the color coding refers to the temperature).
The relation covers cloud dust (yellow), filament dust (orange) and core
dust (green to black). The maximal (minimal) column densities (temperatures)
are far from the values arising from our detailed modeling with a
temperature and density variation along the line-of-sight.
The full range of possible pairs of the two quantities is shown in grey.
This means that neither a single temperature is a good approximation
for lines intersecting a dense core, nor does such an analysis 
produce reliable central column densities or temperatures.

\subsection{Core mass functions, or are they filament mass functions too?}\label{com}
Estimating masses has proven to be a difficult task
for star-forming cores.
The uncertainty in the dust properties and the complex topologies of
the density are just two of the major obstacles encountered.
Correspondingly, core mass functions (CMF) carry large error bars on top of potential
biases from incompleteness.
In this work, we point out two more potential problems namely the inclusion of too much
filament mass, and the break-down of the single temperature approximation that yield
too low masses
and central column densities.
From our analysis, the critical point is not only to limit the mass calculation within smaller radii, 
close to the border between core and filament, but also to avoid adding the contribution of the filament in front
and behind along the line-of-sight. The Abel inversion method based on azimuthally averaged shells or ellipsoids described in R+14 is therefore not well suited to  mass determination for the CMF.

\subsection{Variation of the results when altering the dust properties}\label{dust}
In general, the derived densities scale with the overall shift in FIR 
absorption cross sections
when another dust model is used. This uncertainty remains until we can find out
more about the physical nature of the dust near and in the core.
As a side note we mention here that the surface brightness modeling of the 500
$\mu$m profiles provided better fits when lowering the HGBS opacity by 15 \%. This is not
to be confused with an offset problem of the filter which would add a constant offset
while the offset we find is a factor. For us this may point to
a deviation in the opacity of the dust in L1689B.

The choice of dust opacity becomes even more relevant in the core center where 
coagulation calculations hint at a growth of grains
\citep{1994A&A...291..943O},
extinction measurements indicate a flattening in the extinction law 
\citep[e.g.,][]{2009ApJ...693L..81M},
and scattered MIR radiation seems
to indicate the presence of $\mu$m-sized grains 
\citep{2010A&A...511A...9S}.
The existing {\it Spitzer} images of L1689B
at 3.6\,$\mu$m are difficult to interpret because of the presence of a nearby bright 
star, and the existence of coreshine as scattered light from large grains 
can not be assessed. However, current coreshine measurements in prestellar cores seem to speak against
a strong variation in dust properties on scales of 10 kau 
\citep{2013A&A...559A..60A}.

\subsection{Very cold dust (< 8 K) has a small mass percentage in L1689B}
Our exploration of the
parameter space for the inner shells have not produced solutions where 
dust with temperatures below 8 K is found outside the two inner cells.
 While the contribution of these two shells to the column density are substantial
and cause the remaining spread in our central temperature estimates, their contribution
to the core mass in L1689B is small (< 10 \%) and far from the 30-70 estimated by
\citet{2015A&A...574L...5P}.

This latter work also mentions preliminary modeling of emission line data which requires
central H$_2$ number densities above $\sim$2$\times$10$^{12}$ m$^{-3}$ as indicated
in Fig.~\ref{nT}. The presented range of values for the central shell includes these densities
and the correlation found between density and temperature for this solution branch 
can actually be used in the modeling now. 

\begin{acknowledgements}
We thank A. Roy and P. Palmeirim for providing the Abel transform profile data.
The anonymous referee we thank for a number of constructive comments helping to
improve the manuscript.
\end{acknowledgements}

\bibliographystyle{aa} % style aa.bst
\bibliography{L1689B} % your references Yourfile.bib

\Online

\begin{appendix}
\section{Compulsory acknowledgments}
This research has made use of data from the {\it Herschel} Gould Belt survey (HGBS) project (http://gouldbelt-herschel.cea.fr). The HGBS is a {\it Herschel} Key Programme jointly carried out by SPIRE Specialist Astronomy Group 3 (SAG 3), scientists of several institutes in the PACS Consortium (CEA Saclay, INAF-IFSI Rome and INAF-Arcetri, KU Leuven, MPIA Heidelberg), and scientists of the {\it Herschel} Science Center (HSC). 
{\it Herschel} is an ESA space observatory with science instruments provided by European-led Principal Investigator consortia and with important participation from NASA. SPIRE has been developed by a consortium of institutes led by Cardiff Univ. (UK) and including Univ. Lethbridge (Canada); NAOC (China); CEA, LAM (France); IFSI, Univ. Padua (Italy); IAC (Spain); Stockholm Observatory (Sweden); Imperial College London, RAL, UCL-MSSL, UKATC, Univ. Sussex (UK); Caltech, JPL, NHSC, Univ. Colorado (USA). This development has been supported by national funding agencies: CSA (Canada); NAOC (China); CEA, CNES, CNRS (France); ASI (Italy); MCINN (Spain); SNSB (Sweden); STFC (UK); and NASA (USA). PACS has been developed by a consortium of institutes led by MPE (Germany) and including UVIE (Austria); KUL, CSL, IMEC (Belgium); CEA, OAMP (France); MPIA (Germany); IFSI, OAP/AOT, OAA/CAISMI, LENS, SISSA (Italy); IAC (Spain). This development has been supported by the funding agencies BMVIT (Austria), ESA-PRODEX (Belgium), CEA/CNES (France), DLR (Germany), ASI (Italy), and CICT/MCT (Spain). The James Clerk Maxwell Telescope has historically been operated by the Joint Astronomy Centre on behalf of the Science and Technology Facilities Council of the United Kingdom, the National Research Council of Canada and the Netherlands Organisation for Scientific Research. Additional funds for the construction of SCUBA-2 were provided by the Canada Foundation for Innovation.  IRAM is supported by INSU/CNRS (France), MPG (Germany) and IGN (Spain). This research has made use of NASA's Astrophysics Data System.

\end{appendix}

\end{document}